\documentclass{aastex7}

\usepackage{color} 
\usepackage{amsmath}
\usepackage{bm}
\usepackage{mathtools}

\newcommand{\tred}[1]{#1}
\newcommand{\figdir}{.}

\begin{document}

\title{
  What determines the brightness of magnetically open solar corona?: \\
  Insights from three-dimensional radiative MHD simulations and observations
}


\author[0000-0002-1007-181X]{Haruhisa Iijima}
\affiliation{
  Institute for Space-Earth Environmental Research, Nagoya University, Furocho, Chikusa-ku, Nagoya, Aichi 464-8601, Japan
}
\affiliation{
  Institute for Advanced Research, Nagoya University, Furocho, Chikusa-ku, Nagoya, Aichi 464-8601, Japan
}
\email{h.iijima@isee.nagoya-u.ac.jp}


\begin{abstract}
  We investigate the relationship between solar coronal holes and open-field regions using three-dimensional radiative magnetohydrodynamic (MHD) simulations combined with remote-sensing observations from the Solar Dynamics Observatory (SDO). Our numerical simulations reveal that magnetically open regions in the corona can exhibit brightness comparable to quiet regions, challenging the conventional view that open-field regions are inherently dark coronal holes. We find that the coronal brightness is primarily determined by the total energy input from photospheric magnetic activities, such as the small-scale dynamo, rather than differences in dissipative processes within the corona. Using synthesized EUV intensity maps, we show that brightness thresholds commonly used to identify coronal holes may overlook open-field regions, especially at lower spatial resolutions. Observational analysis utilizing SDO/HMI and AIA synoptic maps supports our simulation results, demonstrating that magnetic field extrapolation techniques, such as the Potential Field Source Surface (PFSS) model, are sensitive to the chosen parameters, including the source surface height. We suggest that discrepancies in estimates of open magnetic flux (the ``open flux problem'') arise both from the modeling assumptions in coronal magnetic field extrapolation and systematic biases in solar surface magnetic field observations. Our findings indicate the need for reconsidering criteria used to identify coronal holes as indicators of open-field regions to better characterize the solar open magnetic flux.
\end{abstract}


\section{Introduction\label{sec:intro}}

Coronal holes are dark, cool regions in the solar corona. The formation mechanism of these dark coronal holes has been discussed for many years. One popular model is that the brightness of a coronal hole is determined by its magnetic connectivity to interplanetary space \citep{pneuman1973solar}. If the coronal magnetic field is open to space, the mechanical energy input is used to both accelerate the solar wind and heat the coronal plasma. Therefore, the coronal brightness becomes smaller than in magnetically closed regions. From this concept, the term ``coronal hole'' is often used to refer to the magnetically open region on the solar surface, independent on the coronal brightness.

If most magnetic flux in the interplanetary space comes from coronal holes, the total flux from coronal holes should be equal to the interplanetary magnetic flux as estimated from the in-situ observations. It is known that the strength of the interplanetary magnetic field does not show a strong dependence on the solar latitude \citep[e.g., ][]{smith1995ulysses}. Thus, the total flux of open magnetic field can be estimated from several methods: (1) measuring the solar surface magnetic flux in coronal holes, (2) using the extrapolation method of coronal magnetic field, and (3) using the in-situ observation of interplanetary magnetic field. However, a discrepancy between these values has been reported by several groups \citep{wang1995solar,hayashi2003mhd,riley2007alternative,yeates2010nonpotential,shiota2016magnetohydrodynamic,linker2017open}.~\cite{linker2017open} named this discrepancy the ``open flux problem''.

There are several potential solutions to solve the ``open flux problem''.
One possibility is a systematic error in the polarimetric magnetic field observations on the solar surface \citep[e.g.][]{riley2019can,wang2022magnetograph}.
Another possibility is a systematic bias in the coronal magnetic field extrapolation.
One example is the potential field source surface (PFSS) model \citep{schatten1969model}, which assumes that the magnetic field is open above a certain height.
However, the height of the source surface, where the magnetic field is assumed to be open, is not well constrained.
The source surface height is often set to \(2.5 R_\odot\) \citep{schatten1969model,hoeksema1983structure}, validated by the correspondence between open-field regions and coronal holes \citep{riley2019can}.
However, the other (lower) values for the source surface height have been suggested to better explain other metrics, including the interplanetary magnetic field strength, {magnetic field direction inferred from polarimetric measurements, open-field source identified from temperature-dependent emission structures,} magnetic polarity reversals, streamer belts, or solar wind velocity \citep[e.g.][]{levine1982open,lee2011coronal,habbal2001predominance,habbal2021identifying,badman2021measurement,badman2022constraining,huang2024adjusting,tokumaru2024coronal}.

In this study, we investigate the formation mechanism of coronal holes based on two approaches: first, three-dimensional radiative magnetohydrodynamic simulations of a realistic magnetically open solar corona, and second, remote-sensing observations of solar coronal holes using the PFSS model.
To avoid the confusion, we define coronal holes as regions with weak intensity of X-rays or extreme ultraviolet, and strictly distinguish {these} dark regions from open-field {regions}, where magnetic field lines are open to interplanetary space.
In Sec.~\ref{sec:simulation}, we show a series of three-dimensional simulations of a magnetically open corona, including coronal holes. We find that magnetically open corona can be bright, similar to quiet regions outside coronal holes. In Sec.~\ref{sec:observation}, we use the PFSS model to extrapolate the coronal magnetic field and observations by the Solar Dynamics Observatory (SDO) to further examine the relationship between coronal holes and open-field regions. Sec.~\ref{sec:summary} summarizes the results and discusses the implications of our findings.

\section{Numerical simulations of magnetically open solar corona\label{sec:simulation}}

First, we present a series of three-dimensional magnetohydrodynamic simulations that mimic the heating process of the magnetically open ambient corona, including coronal holes. In observational data, it is very difficult to strictly identify open-field regions independent of the magnetic field extrapolation models. Thus, it is advantageous to investigate the heating process of magnetically open corona using numerical simulations.

\subsection{Basic equations and implementations\label{sec:simulation:eqs}}

We used the radiative MHD code RAMENS \citep[][]{iijima2015effect,iijima2017three,iijima2023comprehensive} to solve one-fluid MHD equations including the effects of partial ionization, field-aligned thermal conduction by electrons, and radiative energy exchange. As our model directly resolves the MHD wave propagation from the solar surface to the corona, we do not employ any empirical heating terms except for the numerical diffusion required for the numerical stability.

The model integrates the one-fluid compressible magnetohydrodynamic equations:
\begin{align}
  \frac{\partial{\rho}}{\partial t}
  + \nabla \cdot \left( \rho \bm{V} \right) &= 0 ,\\
  \frac{\partial \left( {\rho} \bm{V} \right)}{\partial t}
  + \nabla \cdot \left( \rho \bm{V} \bm{V} + P \right)
  &= \frac{1}{4\pi} \left( \nabla \times \bm{B} \right) \times \bm{B}
  + \rho\bm{g} ,\\
  \frac{\partial e_{\rm int} }{\partial t}
  + \nabla \cdot \left( e_{\rm int} \bm{V} \right)
  &= - P \nabla\cdot\bm{V}
  + Q_{\rm vis} + Q_{\rm res} + Q_{\rm rad} + Q_{\rm cnd} ,\\
  \frac{\partial\bm{B}}{\partial t}
  + \nabla \times \left( \bm{V} \times \bm{B} \right)
  &= 0 ,
\end{align}
where \({\rho}\), \(P\), \(e_{\rm int}\), \(\bm{V}\), \(\bm{B}\), and \(\bm{g}\) are mass density, gas pressure, internal energy density, velocity field, magnetic field, and gravitational acceleration by the Sun, respectively.
\(Q_{\rm vis}\) and \(Q_{\rm res}\) indicate viscous and resistive heating rates, respectively, caused by the numerical diffusion in the equations of motion and induction.
Although we do not solve the equation of the total energy conservation explicitly, the energy-consistent discretization keeps the total energy conservation throughout the time integration \citep{iijima2021energy}.

We assumed the local thermodynamic equilibrium (LTE) in the equation of state. The latent heat by the partial ionization transports two-thirds of the total convective energy in the near-surface convection zone \citep{stein1998simulations}. Thus, we used a tabulated equation of state accounting for the latent heat change including the most abundant twelve elements using the partition function by \cite{irwin1981design}. The latent heat of the molecular hydrogen is also taken into account.

The radiative heating rate \(Q_{\rm rad}\) is defined as a combination of optically thick and optically thin rates
\begin{align}
  Q_{\rm rad} = (1-f_{\rm thin}) Q_{\rm thick} + f_{\rm thin} Q_{\rm thin},
\end{align}
where
\begin{align}
  f_{\rm thin} = 1 - {\rm e}^{- (\rho / \rho_{\rm thin})^2 }.
\end{align}
Here, the threshold value \(\rho_{\rm thin}\) is set to \(10^{-11}\) g cm\(^{-3}\).

The optically thick radiative heating rate \(Q_{\rm thick}\) is derived by integrating the radiative transfer equation in the gray approximation \citep{iijima2017three}. We used the LTE Rosseland mean opacity from OPAL project \citep{iglesias1996updated} combined with the low-temperature opacity by \citep{ferguson2005low}. The RAMENS has a capability to switch between 3D and 1D (plane-parallel) radiative transfer equations. In this study, we used the 1D version as the difference is minor in the spatial resolution used here \citep{hotta2019weak}.

The optically thin radiative heating rate \(Q_{\rm thin}\) is defined by
\begin{align}
  Q_{\rm thin} = - n_e n_H \Lambda(T) ,
\end{align}
where \(n_e\) and \(n_H\) are the number densities of electron and hydrogen nuclei, respectively. The radiative loss function \(\Lambda(T)\) is calculated from the Chianti atomic database \citep{dere2019chianti}.

The effect of field-aligned thermal conduction is included as a conductive heating rate
\begin{align}
  Q_{\rm cnd} = \nabla \left[
    \sigma_{\rm SH} T^{5/2} \bm{b}
    \left( \bm{b} \cdot \nabla \right) T
  \right] ,
\end{align}
where \(\sigma_{\rm SH}\) is set to \(10^6\) in c.g.s.\ units to mimic thermal conductivity by electrons {and \(\bm{b}\) is the unit vector in the magnetic field direction}. The transition region between the chromosphere and corona becomes very thin during the time evolution. In this study, we broaden the transition region numerically by enhancing the thermal conduction coefficient while keeping the coronal energy balance unchanged \citep{iijima2021new}.

\subsection{Simulation setup\label{sec:simulation:setup}}

We performed a series of three-dimensional simulations of the magnetically open corona with different magnetic field parameters. The vertical domain spans from \(15\) Mm below to \(222\) Mm above the solar surface. The horizontal domain size at the solar surface is \(48\times48\) Mm\(^2\). The number of grid points is \(256 \times 256 \times 960\) in the \(x\), \(y\), and \(z\) directions, respectively. The grid size is \(187.5\) km in the horizontal direction and \(48.2\) km in the vertical direction at the solar surface. We assumed non-uniform grid spacing in the vertical direction (\(z\)-direction).

The basic equations are solved on the orthogonal coordinates identical to~\cite{matsumoto2021full}. The periodic boundary conditions are assumed in both horizontal directions. While the small-scale dynamo action spontaneously generates photospheric magnetic structure in our simulations, the vertical magnetic flux is conserved during the time integration due to the horizontal periodic boundary conditions. Following~\cite{kopp1976dynamics}, we incorporated the super-radial expansion of magnetic flux (or horizontal size of numerical domain) near \(z \sim 70\) Mm, so that the interplanetary magnetic field strength is kept constant (between \(0.99-1.17\) G at \(z = 222\) Mm) in all cases.

We set a semi-transparent top boundary condition to mimic the energetics in the ambient magnetically open corona. To reduce the computational cost, we did not included the solar wind acceleration region in contrast to~\cite{iijima2023comprehensive}. At the top boundary, the outward Alfv\'en waves can freely escape from the numerical domain. The amplitude of inward Alfvén waves is set to zero. The outward/inward Alfv\'en waves are approximated by the horizontal components of the Els\"asser variables. Similarly, the top boundary is open for upward flow and closed for downward flow. The thermal conduction flux is set to zero. This simulation setup will overestimate the outgoing energy flows (especially the Poynting flux) as the reflected Alfv\'en waves and conductive flux (by the dissipation in upper layer) should exist in the real solar corona. Thus, we believe that the current configuration can mimic a magnetically open corona darker than the real Sun, in terms of the top boundary condition.

The bottom boundary is located in the solar convection zone. The velocity field is set nearly symmetric so that the plasma can flow in and out of the computational domain. The total (gas plus magnetic) pressure is kept constant at the bottom boundary. Here, we allowed weak deviation of total pressure from the prescribed value to avoid the over-excitation of acoustic waves in the simulated convection zone. The specific entropy of the downflow is set symmetric, whereas the specific entropy is set to a constant value in the upward flow at the bottom boundary.

In this study, we select two parameters to control the coronal heating process: the magnetic flux imbalance \(B^z_{\rm ave}\) and the small-scale dynamo parameter \(B^h_{\rm in}\). The magnetic flux imbalance is defined as the signed average of vertical magnetic flux density at the surface \(z = 0\)
\begin{equation}
  B^z_{\rm ave} = \left<B_z(z = 0)\right>_{x,y} = {\rm const.},
\end{equation}
where the average \(\left<.\right>_{x,y}\) is taken over the whole horizontal domain.
As we impose the periodic boundary condition in the horizontal direction, the signed average of vertical magnetic flux density is conserved throughout the time evolution. We set the parameter \(B^z_{\rm ave}\) to values of \(3\), \(6\), \(12\), and \(24\) G. {The later values (12 and 24 G) are larger than the representitive values of coronal holes. We use these values because (1) to investigate the effect of magnetic flux imbalance on the coronal brightness and (2) to cover the parameter ambiguity potentially caused by the magnetic field measurements in the photosphere \citep[][also see a discussion in Sec.~\ref{sec:summary:ss_height}]{milic2024spatial,sinjan2024magnetograms}.}

Another control parameter is the strength of the simulated small-scale dynamo.~\cite{rempel2014numerical} reported that the magnetic energy in the inflow region at the bottom boundary determines the realized strength of the simulated quiet Sun magnetism. To mimic this approach, we set the maximum value on the horizontal magnetic field strength in the inflow region at the bottom boundary as
\begin{align}
  B_{x, y}(z=z_{\rm bot}) = \max\left[
    -B_h^{\rm in}, \min\left[
      B_h^{\rm in}, B_{x, y}^{\rm sym}(z=z_{\rm bot})
    \right]
  \right],
\end{align}
where \(B_{x, y}^{\rm sym}\) is the \(x\) and \(y\) components of the horizontal magnetic field extrapolated into the boundary region by assuming zero vertical gradient. We set \(B_h^{\rm in}\) to \(0\) (minimum), \(710\) G, \(1760\) G, and \({\infty}\) (maximum). More precisely, we set the horizontal magnetic field to be anti-symmetric when \(B_h^{\rm in}=0\) to suppress the dynamo action effectively.

Each simulation is relaxed to a quasi-steady corona through multiple stages. The initial condition is the one-dimensional stratification with horizontal uniform small perturbations to the mass density. The upper boundary was set to 900 km above the solar surface for the first 5 days. During this period, the magnetic field was gradually amplified by the simulated dynamo action and a quasi-steady state was reached. We then extrapolated the corona assuming the potential field extrapolation and hydrostatic equilibrium. The coronal simulations were relaxed for 6 h. In this study, only the last 2 h of data are analyzed.

\subsection{Overall structure of simulated corona\label{sec:simulation:overall}}

\begin{figure}[ht!]
  \epsscale{0.4}
  \plotone{{\figdir}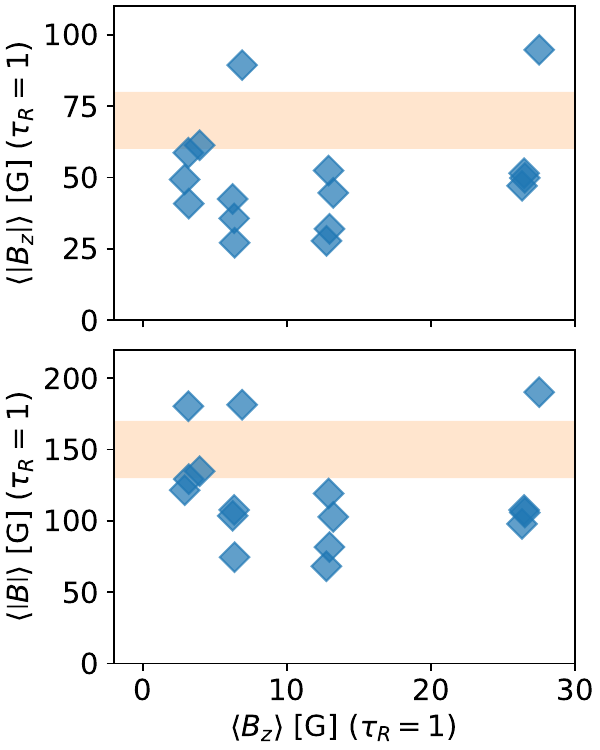}
  \caption{
    Realized strength of small-scale dynamo measured by \(\left<|B_z|\right>\) and \(\left<|B|\right>\) as a function of the magnetic flux imbalance \(\left<Bz\right>\).
    All quantities are evaluated at the Rosseland-mean optical depth unity (\(\tau_R=1\)).
    Each point corresponds to a different simulation case with different values of \(B_h^{\rm in}\) and \(B^z_{\rm ave}\).
    Orange shaded regions indicate typical ranges in the quiet Sun.
  }\label{fig:sim:bsrf}
\end{figure}

{
Figure~\ref{fig:sim:bsrf} shows the strength of the small-scale dynamo realized in our simulations.
We measured the root-mean-square of the vertical magnetic field \(\left<|B_z|\right>\) and the total magnetic field strength \(\left<|B|\right>\) at the Rosseland-mean optical depth unity (\(\tau_R=1\)) as a measure of the small-scale dynamo strength.
The realized magnetic field strengths are weakly dependent on the magnetic flux imbalance \(\left<B^z_{\rm ave}\right>\).
The vertical dispersion of data points is caused by the different values of the inflow magnetic field strength \(B_h^{\rm in}\) with a randomness of the thermal convection.
Orange shaded regions indicate typical ranges in the quiet Sun suggested by comparative studies between radiative MHD simulations and observations \citep{rempel2014numerical,danilovic2016internetwork,del2018novel,rempel2023small}.
Our simulations fully cover the lower and upper ranges of the small-scale dynamo strength outside active regions.
We note that the proxies of small-scale magnetic field strength, \(\left<|B_z|\right>\) and \(\left<|B|\right>\), strongly depend on the spatial resolution.
Considering moderately horizontal grid size in our simulations, we speculate that the real Sun lies around \(B_h^{\rm in} \approx 710\text{--}1760\) G.
}

\begin{figure}[ht!]
  \plotone{{\figdir}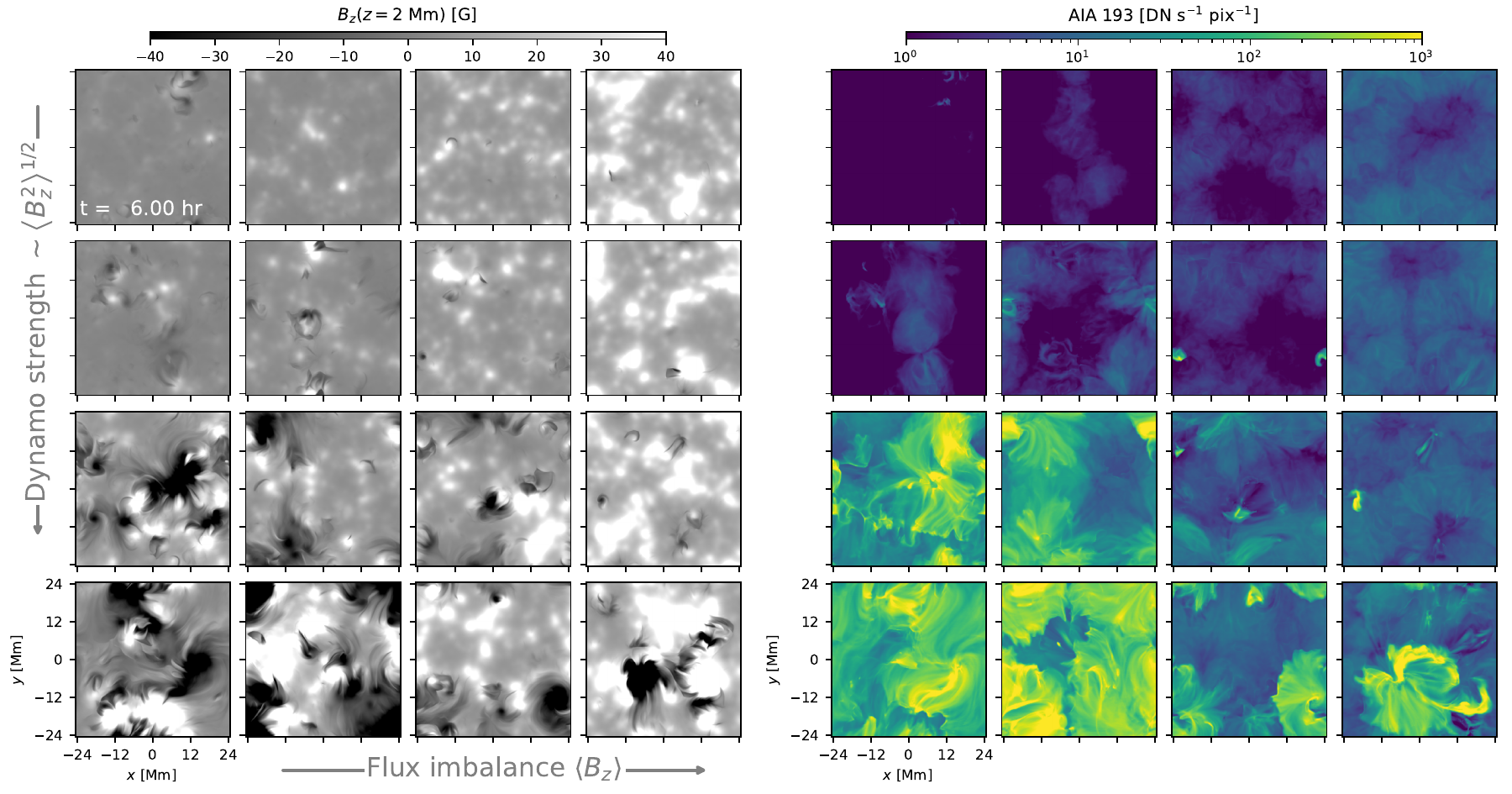}
  \caption{
    Overall structure of the simulated solar corona.
    Shown are vertical component of magnetic field at the coronal base (2 Mm above the surface; left) and synthesized AIA 193 \AA\ intensity.
    Each sub-panel corresponds to a different simulation case with different values of \(B_h^{\rm in}\) and \(B^z_{\rm ave}\).
    From top to bottom, the inflow magnetic field strength \(B_h^{\rm in}\) is set to \(0\), \(710\) G, \(1760\) G, and \({\infty}\), respectively.
    From left to right, the magnetic flux imbalance \(B^z_{\rm ave}\) is set to \(3\), \(6\), \(12\), and \(24\) G, respectively.
    {The animated version of this figure shows time evolution of the simulated corona over 30 minutes of physical time.}
  }\label{fig:sim:overview}
\end{figure}

Figure~\ref{fig:sim:overview} shows the overall structure of the simulated solar corona in all 16 cases with different values of \(B_h^{\rm in}\) and \(B^z_{\rm ave}\). The left panel shows the vertical component of the magnetic field at the coronal base (2 Mm above the surface). The magnetic structure becomes more unipolar with a stronger flux imbalance and weaker small-scale dynamo. The negative sign of vertical magnetic field at the coronal height is a measure of closed coronal loops, which are more dominant in the case of stronger small-scale dynamo. The right panel shows the synthesized AIA 193 \AA\ intensity. The synthesized AIA 193 \AA\ intensity is calculated from the coronal temperature and density using the CHIANTI atomic database \citep{dere2019chianti}. The AIA 193 \AA\ intensity has been a popular measure of coronal holes \citep[e.g.,][]{linker2017open}. Our simulations show that the magnetically open corona is not always dark. With a strong small-scale dynamo, the EUV brightness of simulated corona can be high, similar to quiet regions. The brightness of coronal holes is not only determined by their degree of unipolarity but also by the small-scale dynamo. The magnetic topology (open or closed) is still important, but the small-scale dynamo also plays a significant role in determining the brightness of the magnetically open solar corona.

\begin{figure}[ht!]
  \epsscale{0.8}
  \plotone{{\figdir}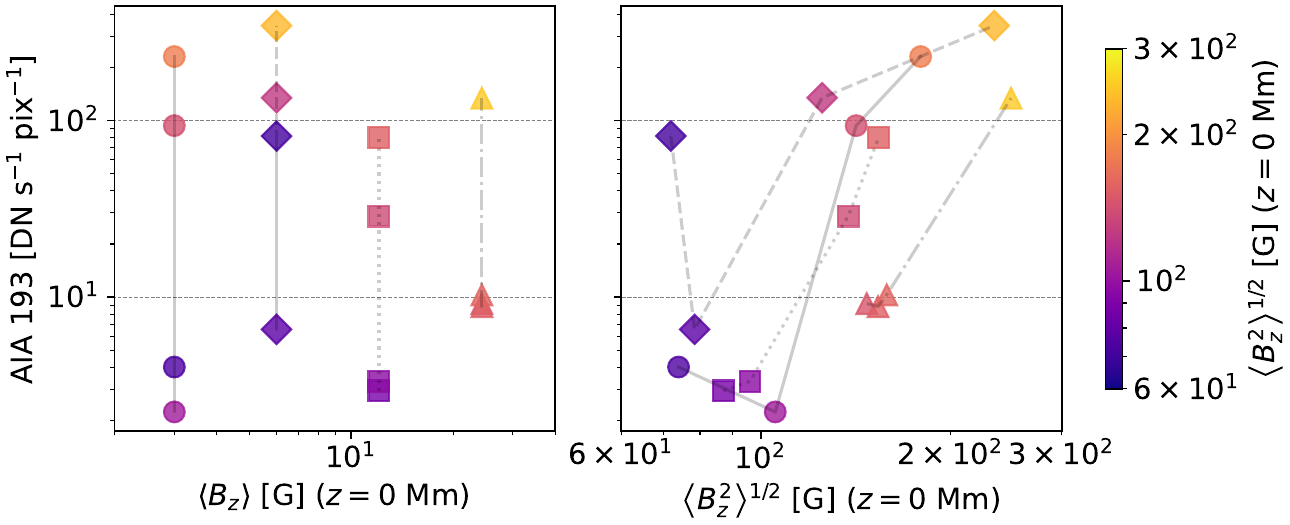}
  \caption{
    Dependence of synthesized AIA 193 \AA\ intensity on the vertical magnetic flux imbalance (left) and the root-mean-square of the vertical magnetic field (right).
    The average is taken over the whole horizontal domain and the last 2 h of each simulation.
    Two thin gray horizontal lines indicate the typical intensity of the quiet Sun (top) and the coronal hole (bottom).
  }\label{fig:sim:aia_cmp}
\end{figure}

Figure~\ref{fig:sim:aia_cmp} shows the dependence of the synthesized AIA 193 \AA\ intensity on the vertical magnetic flux imbalance (left) and the root-mean-square of the vertical magnetic field (right). The average is taken over the whole horizontal domain and the last 2 h of each simulation. The synthesized AIA 193 \AA\ intensity shows a clear positive correlation with the root-mean-square of the vertical magnetic field, but a weak correlation with the vertical magnetic flux imbalance. This suggests that, on average (or at low spatial resolutions), the brightness of magnetically open corona is strongly affected by the small-scale dynamo and weakly affected by the magnetic flux imbalance.

\begin{figure}[ht!]
  \epsscale{0.8}
  \plotone{{\figdir}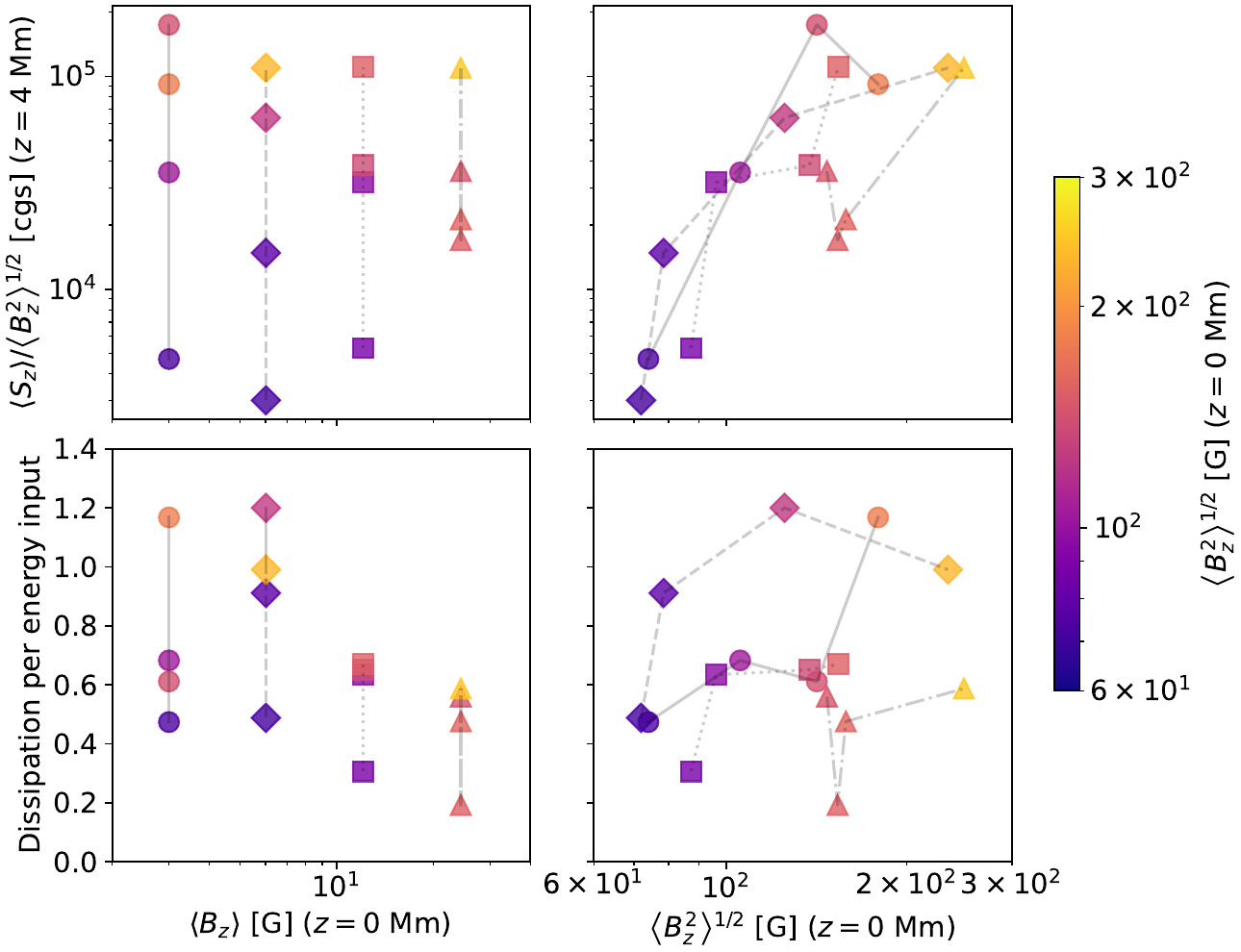}
  \caption{
    Energy input (top) and dissipation rate (bottom) in the simulated solar corona.
    Shown are the dependence on the vertical magnetic flux imbalance (left) and the root-mean-square of the vertical magnetic field (right).
    The average is taken over the whole horizontal domain and the last 2 h of each simulation.
  }\label{fig:sim:energy_input}
\end{figure}

The plasma heating process in the simulated corona is summarized in Figure~\ref{fig:sim:energy_input}. We measured the horizontally and temporally averaged upward Poynting flux at the coronal base (at \(z = 4\) Mm) as a measure of energy input to the simulated corona. We selected the Poynting flux divided by the magnetic field strength \(\left<S_z\right>/\left<B_z^2\right>^{1/2}\) at \(z = 4\) Mm as the vertical axis of the top panels. In previous studies of coronal heating modeling, the input energy flux is assumed to be nearly proportional to the local magnetic field strength \citep[e.g.,][]{sokolov2013magnetohydrodynamic,wang2020small} at the coronal foot point. This assumption of proportionality between the Poynting flux and the magnetic field strength is consistent with the excitation and propagation of Alfv\'en waves in the magnetized solar atmosphere \citep[e.g.,][]{hollweg1984alfvenic,cranmer2011testing,shoda2021corona}. In our simulations, \(\left<S_z\right>/\left<B_z^2\right>^{1/2}\) shows almost no correlation to the magnetic flux imbalance \(\left<B_z\right>\) with large dispersion. This dispersion is caused by the positive correlation between \(\left<S_z\right>/\left<B_z^2\right>^{1/2}\) and \(\left<B_z^2(z = 0)\right>^{1/2}\) (top right panel). The photospheric magnetic energy \(\left<B_z^2(z = 0)\right>^{1/2}\) is a measure of the small-scale dynamo strength.~\cite{iijima2023comprehensive} argues that the interchange reconnection (that strongly depends on the photospheric magnetic energy) provides a significant amount of energy flux into the corona and solar wind. We speculate that this positive correlation between \(\left<S_z\right>/\left<B_z^2\right>^{1/2}\) and \(\left<B_z^2(z = 0)\right>^{1/2}\) is also caused by the interchange magnetic reconnection in the chromosphere and low corona.

We evaluated the total energy dissipation in the simulated corona as a sum of viscous and resistive heating rates between \(z = 4\) Mm and \(z = 100\) Mm
\begin{equation}
  \epsilon_{\rm diss}
  = \int_{z = 4\ {\rm Mm}}^{100\ {\rm Mm}}
    \frac{\left<Q_{\rm vis} + Q_{\rm res} \right>_{x,y,t}}
    {\left<B_z\right>_{x,y,t}}dz,
\end{equation}
where the denominator \(\left<B_z\right>\) is used to take into account the horizontal expansion of numerical domain along the \(z\)-direction.
We choose the upper limit of the integral to be \(100\) Mm to avoid the influence of the top boundary condition. Note that the magnetic field structure is nearly uniformly vertical at this height.
The evaluation of viscous and resistive heating rates is described in~\cite{iijima2021energy}..
As the amount of energy dissipation varies significantly depending on the simulation parameters, we normalized the energy dissipation rate \(\epsilon_{\rm diss}\) by the Poynting flux at the coronal base \(z = 4\) Mm as
\begin{equation}
  r_{\rm diss}
  = \frac{\epsilon_{\rm diss}}
  {\left<S_z(z = 4\ {\rm Mm})\right>_{x,y,t} / \left<B_z(z = 4\ {\rm Mm})\right>_{x,y,t}},
\end{equation}
where \(r_{\rm diss}\) is a non-dimensional parameter that represents the rate of energy dissipation per unit energy input in the simulated corona.
The bottom panels of Figure~\ref{fig:sim:energy_input} show \(r_{\rm diss}\) as a function of the vertical magnetic flux imbalance (left) and the root-mean-square of the vertical magnetic field (right) on the solar surface. In contrast to the Poynting flux (top panels), the dissipation per energy input exhibits a clear negative correlation with the vertical magnetic flux imbalance, suggesting that magnetically unipolar regions are less dissipative than the non-unipolar (magnetically balanced) regions. However, we emphasize that the variation of energy input is much larger (top panels) and will be more important than the dissipation process for the coronal brightness.

\subsection{Dark corona as an indicator of open regions\label{sec:simulation:dark_corona}}

To reveal the correspondence between coronal holes and open-field regions, we define the magnetically open pixels by tracing the magnetic field lines. Starting from an altitude of \(z = 100\) Mm, field lines are traced both down to the photosphere (\(z = 0\) Mm) and up to the top boundary (\(z = 222\) Mm). A pixel is considered magnetically open, thus belonging to an open-field region, if the field line originating from the pixel connects to the top boundary.

We define that a pixel belongs to a coronal hole if the synthesized AIA 193 \AA\ intensity is lower than a threshold value. This threshold value can be chosen by hand. In most cases in this paper, we use the threshold value of \(20\) DN s\(^{-1}\) pix\(^{-1}\), as a typical brightness that distinguishes coronal holes from quiet regions.

\begin{figure}[ht!]
  \epsscale{0.8}
  \plotone{{\figdir}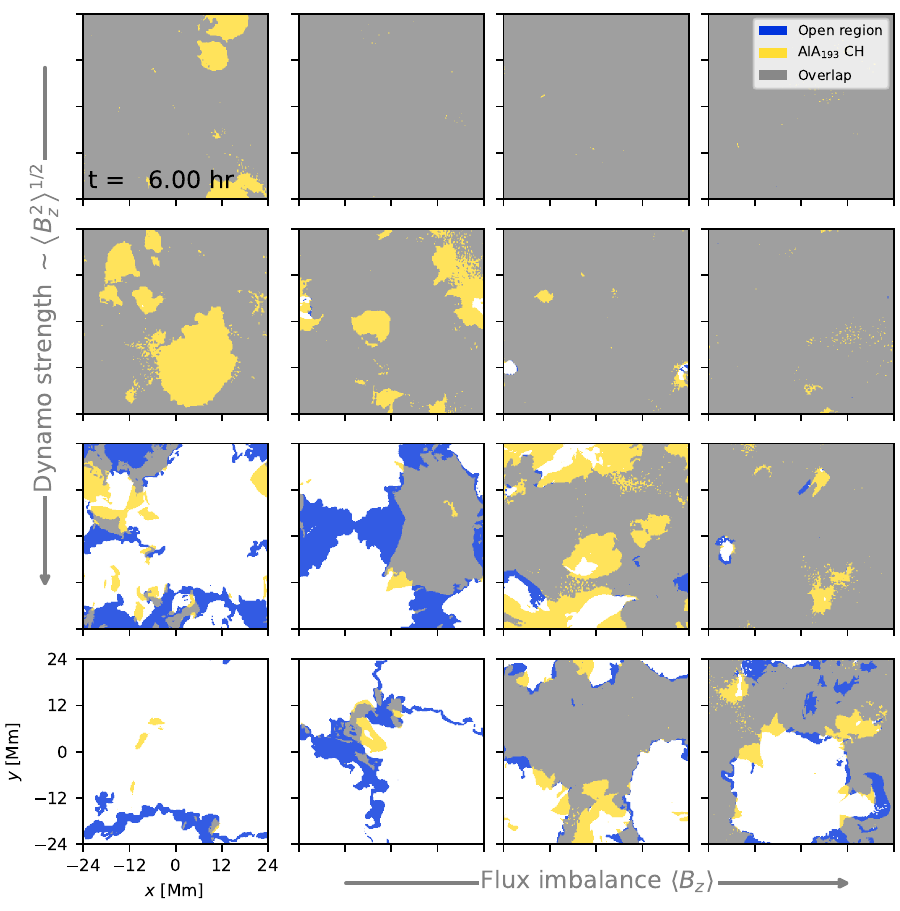}
  \caption{
    Spatial distribution of true open-field region (determined by tracing magnetic field lines) and coronal holes defined by the synthesized AIA 193 \AA\ intensity with a threshold value of \(20\) DN s\(^{-1}\) pix\(^{-1}\).
  }\label{fig:sim:aia_ch_map}
\end{figure}

Figure~\ref{fig:sim:aia_ch_map} shows the spatial distribution of the true open-field region and coronal holes defined by the synthesized AIA 193 \AA\ intensity. Each horizontal pixel is classified as open or closed from the connectivity of the magnetic field lines at an altitude of \(z = 2\) Mm. In general, the coronal brightness (coronal hole) is well correlated with the open-field region. However, there are some pixels that are classified as open but not coronal holes or vice versa. This missclassification is especially significant in the cases of strong small-scale dynamo (bottom rows) or small magnetic flux imbalance (left columns), where the area fraction of open-field region is small. Small open regions are more likely to be misclassified as quiet regions.

\begin{figure}[ht!]
  \epsscale{0.8}
  \plotone{{\figdir}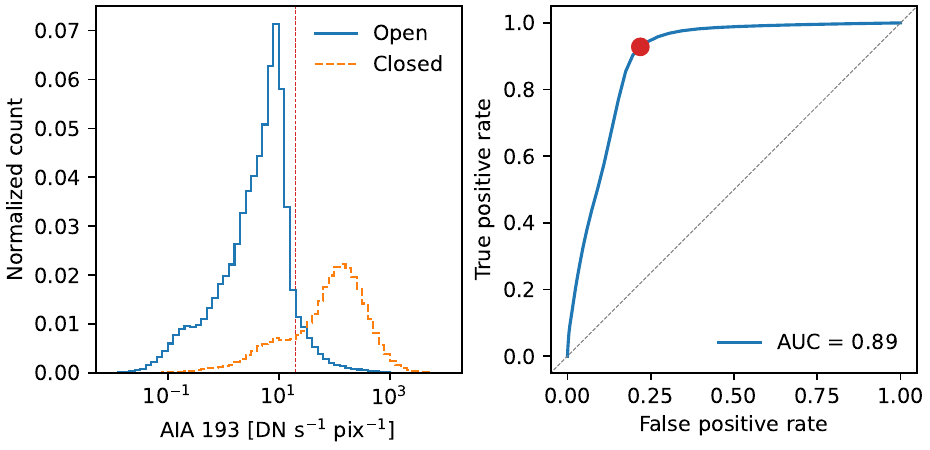}
  \caption{
    Binary classification of open-field regions by the synthesized AIA 193 \AA\ intensity.
    The histograms of the open and closed regions (left) and the ROC curves (right) are shown.
    The estimated AUC is 0.89.
    The optimal intensity threshold by the Youden's J statistic is \(19.9\) DN s\(^{-1}\) pix\(^{-1}\) (shown as a red dashed line in the left panel and a red circle in the right panel).
  }\label{fig:sim:aia_hist}
\end{figure}

For more quantitative analysis, we performed a binary classification of open-field regions by the synthesized AIA 193 \AA\ intensity using receiver operating characteristic (ROC) curve analysis. The results are shown in Figure~\ref{fig:sim:aia_hist}. The histograms of the open and closed regions (left panel) exhibit an overlap between the two distributions. Depending on the threshold value of synthesized AIA 193 \AA\ intensity, the area of open-field regions can be overestimated (false positive) or underestimated (false negative). The synthesized AIA 193 \AA\ intensity is a good indicator of open-field regions, but the threshold value should be carefully chosen.
The right panel shows the ROC curve, which is used to evaluate the performance of a binary classifier. The area under the ROC curve (AUC) is \(0.89\), indicating that the synthesized AIA 193 \AA\ intensity is still a good indicator of open-field regions if the threshold value is carefully chosen. The optimal intensity threshold by the Youden's J statistic is \(19.9\) DN s\(^{-1}\) pix\(^{-1}\), consistent with the typical threshold value (\(20\) DN s\(^{-1}\) pix\(^{-1}\)) used in this paper.

\begin{figure}[ht!]
  \epsscale{0.8}
  \plotone{{\figdir}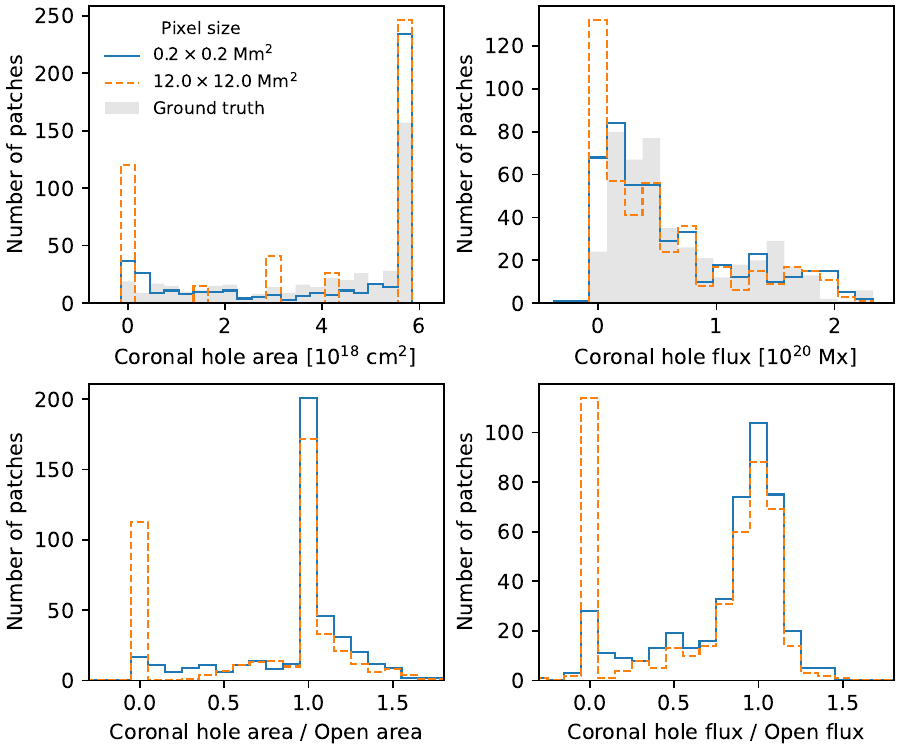}
  \caption{
    Effects of spatial resolution on coronal hole detection.
    All quantities are evaluated within \(24\times24\) Mm\(^2\) horizontal patches.
    (Top left) Histogram of coronal hole area.
    The blue solid line indicates a pixel size of \(0.2\times0.2\) Mm\(^2\) (original resolution), and the orange dashed line indicates a pixel size of \(12\times12\) Mm\(^2\) (rebinned).
    The coronal holes are detected using rebinned maps of the synthesized AIA 193 \AA\ intensity, with a threshold value of \(20\) DN s\(^{-1}\) pix\(^{-1}\).
    \tred{Gray bars show the histogram of the true open area at the original resolution.}
    (Top right) Histogram of the total vertical magnetic flux in coronal holes.
    \tred{Gray bars show the histogram of the true open magnetic flux at the original resolution.}
    (Bottom left) Histogram of the ratio of coronal hole area to the true open area.
    The open pixels are defined by tracing magnetic field lines in the original resolution.
    Patches with no open pixels are excluded.
    (Bottom right) Histogram of the ratio of coronal hole magnetic flux to the true open magnetic flux.
  }\label{fig:sim:ch_vs_open_rebin}
\end{figure}

In observational studies, coronal holes are typically detected in EUV or X-ray images at spatial resolutions lower than those used in our simulations. This lower resolution can mix signals from coronal holes and quiet regions, leading to underestimation of the true coronal hole area and associated ``open'' magnetic flux that would be apparent at higher resolution.

{
To assess the effects of spatial resolution, we rebinned the synthesized intensity to coarser grids and evaluated coronal hole area and magnetic flux at each resolution, while keeping the open/closed classification fixed at the native resolution. The procedure was as follows:
\begin{enumerate}
  \item Trace magnetic field lines in the original high-resolution data to identify open pixels at the coronal base (\(z = 2\) Mm) and compute the true open magnetic flux.
  \item Divide the horizontal domain into \(24\times24\) Mm\(^2\) patches.
  \tred{This yields 448 patches in total (4 per snapshot, 7 snapshots per case with a 20-minute cadence, and 16 cases).}
  \item For each patch, compute the total vertical magnetic flux and the total area of open pixels.
  \item Using the synthesized AIA 193~\AA\ intensity at the given resolution, classify pixels in each patch as coronal hole or not using a threshold of 20~DN~s\(^{-1}\)~pix\(^{-1}\); then compute the corresponding magnetic flux and area.
  \item Repeat step 4 using intensity maps rebinned to coarser pixel sizes by averaging adjacent pixels.
\end{enumerate}
}

{
The effects of spatial resolution on the coronal hole detection are summarized in Figure~\ref{fig:sim:ch_vs_open_rebin}.
The left panels show histograms of coronal hole area detected at different resolutions, compared with the true open area.
\tred{At the rebinned resolution of \(12\times12\) Mm\(^2\), each \(24\times24\) Mm\(^2\) patch contains only \(2\times2\) pixels, so the CH area within a patch can take only five discrete values (0–4 pixels), producing the separated bars seen in the top-left panel (orange line).}
When the coronal hole is detected, the area remains in good agreement with the true open area, regardless of the resolution.
However, at lower resolution, the number of patches without detected coronal holes increases, leading to an underestimation of the total coronal hole area.
The ratio of the total coronal hole area to the total open area is \(1.03\) and \(0.93\) for the original resolution and the rebinned \(12\times12\) Mm\(^2\) maps, respectively.
We note that these values are dependent on the magnetic structures and statistics of the focused region and should not directly be interpreted as the absolute values.
}

{
The right panels in Fig.~\ref{fig:sim:ch_vs_open_rebin} show histograms of the total magnetic flux in coronal holes, compared with the true open magnetic flux.
Similar to the area, the total magnetic flux in coronal holes remains well correlated with the true open magnetic flux when a coronal hole is detected,
although the dispersion of the error between coronal hole flux and true open flux seems to be larger than that of the area.
However, at lower resolutions, the number of patches without detected coronal holes increases, leading to an underestimation of the total coronal hole magnetic flux.
In the patches of our simulations, the ratio of the total coronal hole flux to the total open flux is \(0.91\) and \(0.79\) for the original resolution and the rebinned \(12\times12\) Mm\(^2\) maps, respectively.
The coronal hole magnetic flux shows larger underestimation than the coronal hole area (discussed in previous paragraph) at lower resolutions.
This suggests that a significant fraction of open magnetic flux can effectively be missed or excluded when using lower resolution data, leading to an underestimation of the true total coronal hole area and open magnetic flux.
}

\section{Magnetic property of observed coronal holes\label{sec:observation}}

Our simulations, presented in Sec.~\ref{sec:simulation}, suggested that some magnetically open regions can appear as bright as quiet regions. To test this possibility, we utilized the synoptic maps of the Heliospheric and Magnetic Imager (HMI) and the Atmospheric Imaging Assembly (AIA) from the Solar Dynamics Observatory (SDO).

\subsection{Data and preprocessing\label{sec:observation:data}}

We used the HMI synoptic maps downloaded from \url{http://jsoc.stanford.edu/new/HMI/LOS_Synoptic_charts.html}. Among several variations, we chose \verb|hmi.Synoptic_Mr_720s| that provides the radial component of the magnetic field derived from the line-of-sight magnetograms by dividing by the cosine of the angle from disk center.

The synoptic maps of AIA 193 are downloaded from \url{https://sdo.gsfc.nasa.gov/data/synoptic/}. We removed instrumental degradation based on the degradation curve derived by using the \verb|aiapy| open source software package \citep{Barnes2020}. From all synoptic maps available between June 2010 and September 2021, we removed the maps (1) during flares by applying a threshold to the maximum intensity (i.e., \(5000\) DN s\(^{-1}\) pix\(^{-1}\)), (2) with artificial banded pattern in the latitudinal direction (judged by eyes), (3) with longitudinal missing data (we used maps with missing data near the poles), and (4) between CR2113 and CR2124 (exhibit unnaturally large values of the intensity).

The AIA 193 intensity in the synoptic maps is enhanced near the poles due to the limb brightening effect. To correct this, we followed the method described in \cite{caplan2016synchronic}. The detailed procedure is described in Appendix~\ref{appendix:lbcc}.

In both HMI and AIA synoptic maps, we extrapolated the polar pixels assuming zero latitudinal variation near the poles. After that, we interpolated the AIA synoptic maps into the coordinate of HMI synoptic maps, which uses a cylindrical equal-area (CEA) projection. The AIA synoptic maps are rebinned to match the pixel size of HMI synoptic maps.
This preprocessing resulted in co-aligned HMI and AIA maps with a base pixel size of \(1.2\) Mm, corresponding to the original HMI resolution, which were used for subsequent analysis.

\subsection{Rebinning and magnetic field extrapolation\label{sec:observation:rebining}}

We rebinned the HMI and AIA synoptic maps to lower spatial resolutions by averaging the adjacent pixels in a CEA projection. The typical pixel size of the rebinned maps in this study is approximately \(36\) Mm near the equator, unless otherwise noted.

To identify the open-field regions using the synoptic maps, we extrapolated the coronal magnetic field using the rebinned HMI synoptic maps.
Since identifying open-field regions independently of magnetic field extrapolation models is challenging, we employed a simple Potential Field Source Surface (PFSS) model to delineate the magnetically open regions in the SDO data. To ensure that our conclusions are not sensitive to the details of the coronal field extrapolation method chosen, we examined the influence of the source surface height \(R_{\rm SS}\) by varying this parameter in the PFSS model. The PFSS model we used is the \verb|sunkit-magex| package \citep{Stansby2020}, which is forked from the \verb|pfsspy| package (archived as of August 24, 2023).

\subsection{Correspondence between coronal holes and open-field regions\label{sec:observation:of_vs_ch}}

\begin{figure}[ht!]
  \plotone{{\figdir}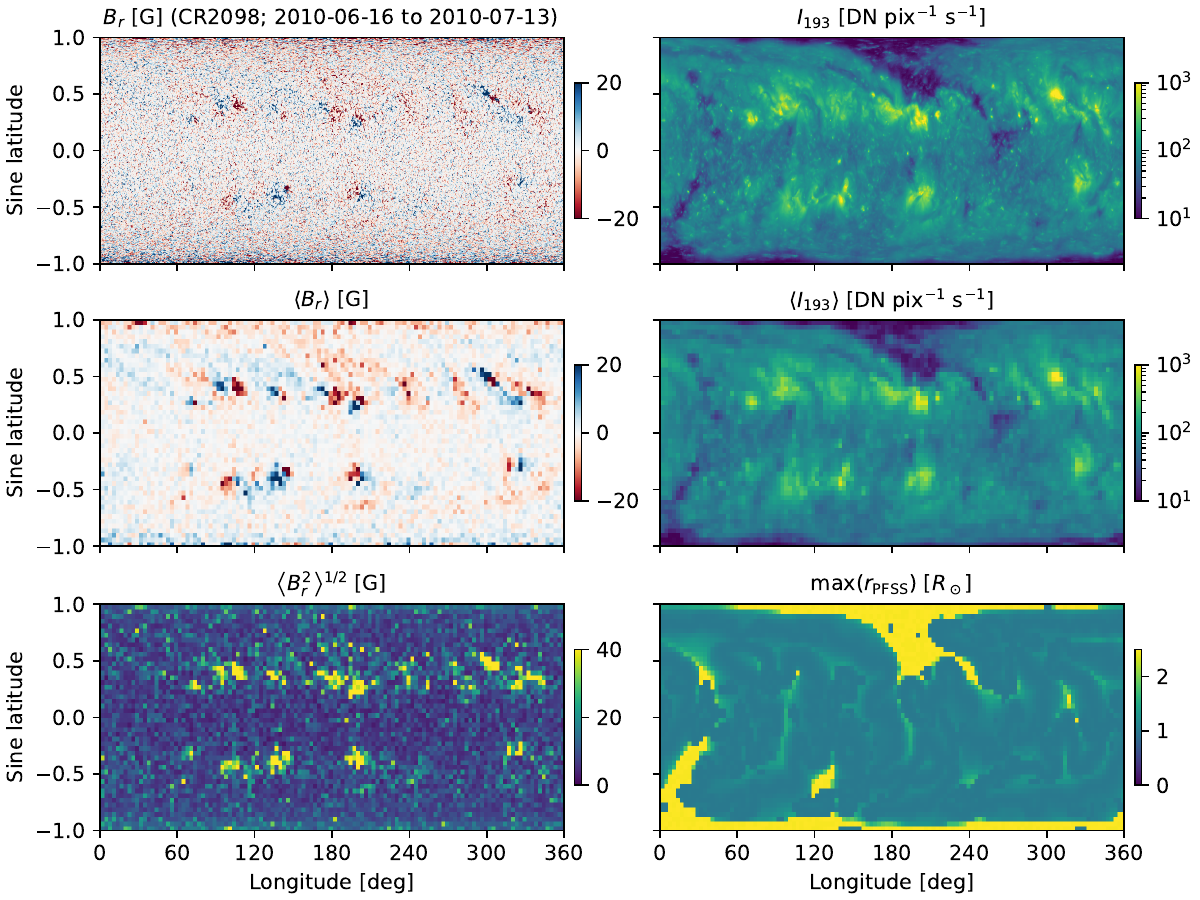}
  \caption{
    HMI (top left) and AIA (top right) synoptic maps for CR2098 after the preprocessing.
    The middle and bottom panels show quantities derived from spatially rebinned data: the mean radial magnetic field \(\left<B_r\right>\) (middle left), the radial magnetic field strength \(\left<B_r^2\right>^{1/2}\) (bottom left), and the mean AIA 193 intensity \(I_{\rm 193}\) (middle right).
    The bottom right panel shows the maximum height of magnetic field lines extrapolated from the rebinned HMI synoptic map (i.e., middle left panel) using the PFSS model with a source surface height \(R_{\rm SS}\) of \(2.5 R_\odot\).
  }\label{fig:obs:maps}
\end{figure}

The resultant maps are shown in Fig.~\ref{fig:obs:maps} for CR2098. The coronal magnetic field is extrapolated with \(R_{\rm SS} = 2.5 R_\odot\), which is a commonly-used value of the source-surface height in PFSS modeling \citep{schatten1969model,hoeksema1983structure}. We evaluated the maximum height of extrapolated magnetic field lines to reduce the artifact from the selection of \(R_{\rm SS}\). The open-field regions identified by the PFSS model show good agreement with the pixels with low AIA 193 intensity, which is a typical indicator of coronal holes.

\begin{figure}[ht!]
  \epsscale{0.9}
  \plotone{{\figdir}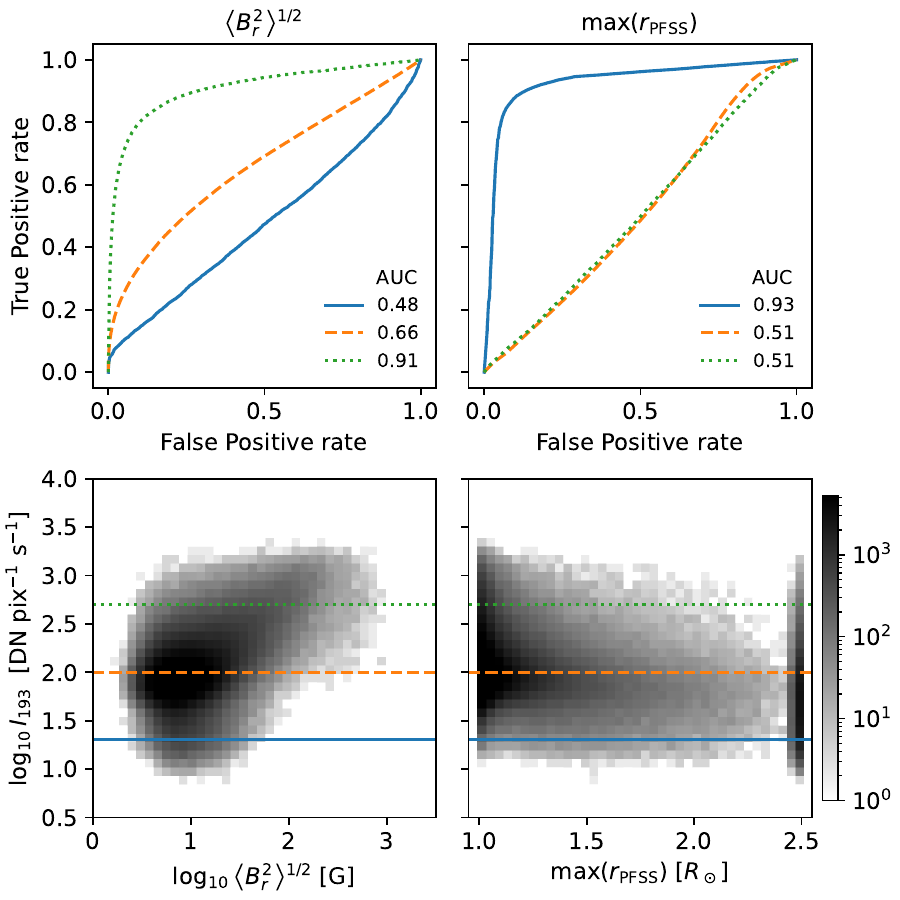}
  \caption{
    Performance of the magnetic field strength on the photosphere (left) and the maximum height of magnetic field lines (right) as a binary classifier of the AIA 193 intensity.
    Top panels show the ROC curves with the area under the curve (AUC) values.
    Three binary classifications of the AIA 193 intensity with 20, 100, 500 DN s\(^{-1}\) pix\(^{-1}\) are shown. These threshold values correspond to the coronal hole, quiet region, and active region, respectively.
    The bottom panels show the two-dimensional probability distribution function of the AIA 193 intensity and each classifier.
  }\label{fig:obs:roc_auc}
\end{figure}

Figure~\ref{fig:obs:roc_auc} shows the performance of the radial magnetic field strength on the photosphere (left) and the maximum height of magnetic field lines (right) as a binary classifier of the AIA 193 intensity. We selected three threshold values of the AIA 193 intensity, 20, 100, 500 DN s\(^{-1}\) pix\(^{-1}\), which correspond to the coronal hole, quiet region, and active region, respectively.
The magnetic field strength on the photosphere is a good indicator of active regions, but not a good indicator of coronal holes. The maximum height of the magnetic field lines is a better indicator of coronal holes, but it cannot identify quiet regions or active regions. This result supports the standard assumption that coronal holes are open-field regions.

\begin{figure}[ht!]
  \epsscale{1.0}
  \plotone{{\figdir}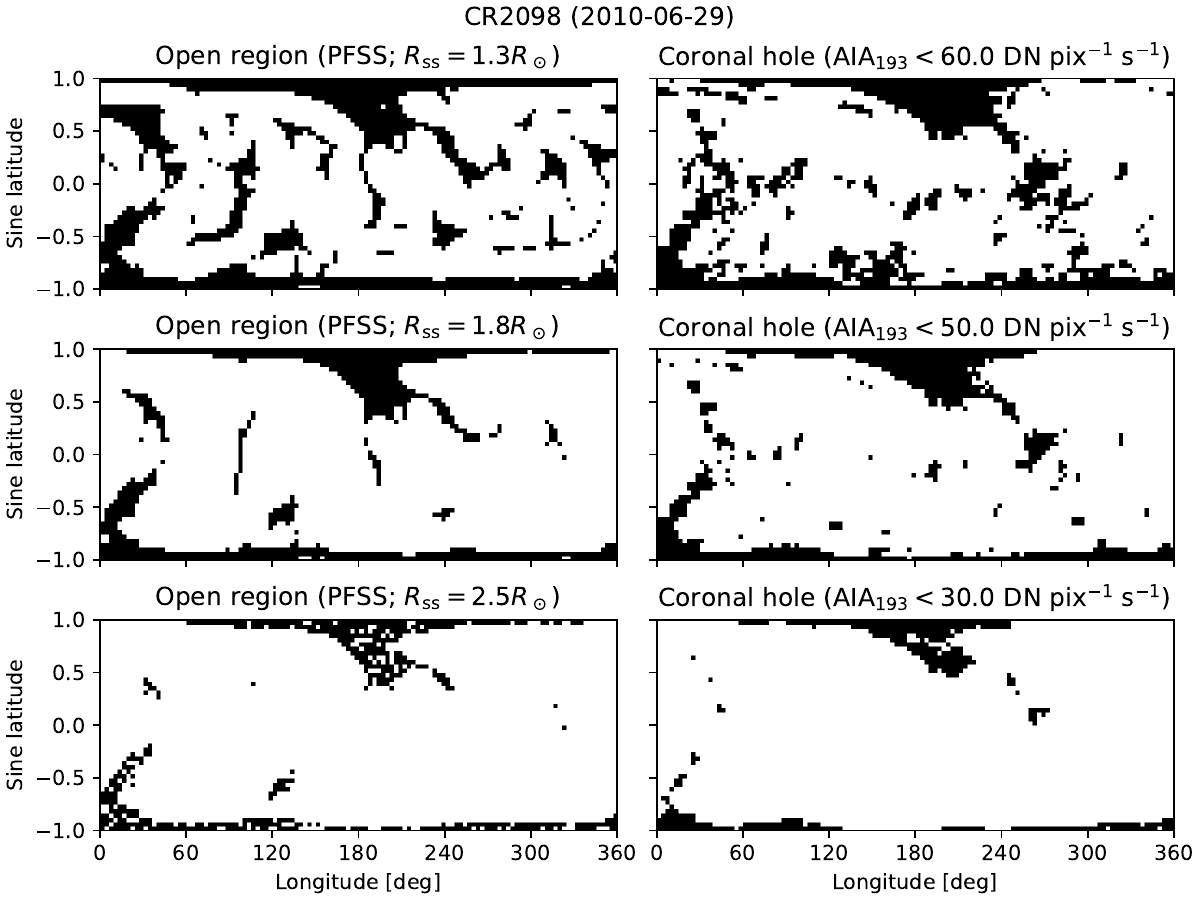}
  \caption{
    The spatial distribution of open-field region (left) and coronal holes (right).
    Left panels show the open-field region defined by the PFSS model with source surface heights of \(1.3 R_\odot\) (top), \(1.8 R_\odot\) (middle) and \(2.5 R_\odot\) (bottom).
    Right panels show the coronal holes defined by the AIA 193 intensity with a threshold value of \(60\) DN s\(^{-1}\) pix\(^{-1}\) (top), \(50\) DN s\(^{-1}\) pix\(^{-1}\) (middle), and \(30\) DN s\(^{-1}\) pix\(^{-1}\) (bottom).
  }\label{fig:obs:of_vs_ch_map}
\end{figure}

On the other hand, we emphasize that the area of coronal holes is strongly dependent on the brightness threshold. Figure~\ref{fig:obs:of_vs_ch_map} shows the spatial correspondence between open-field region and coronal holes. We varied the source surface height \(R_{\rm SS}\) of the PFSS model for detecting open-field regions and the brightness threshold of AIA 193 intensity for detecting coronal holes. Clearly, a lower value of \(R_{\rm SS}\) corresponds to a higher brightness threshold of AIA 193 intensity. As implied from our simulation in Sec.~\ref{sec:simulation}, the brightness of open-field regions can be as high as that of quiet regions, especially if the magnetic flux imbalance is small and spatial resolution is low. Thus, changing the definition of ``coronal holes'' as the indicator of open-field regions can lead to a significant change in the estimated open magnetic flux.

\subsection{Total magnetic flux in coronal holes above specific latitudes\label{sec:observation:ch_flux}}

The discussion in Sec.~\ref{sec:observation:of_vs_ch} depends on the specific modeling of coronal magnetic field, in which several assumptions (e.g. zero electric current, spherically symmetric source surface height) are made. Here, we use the longitudinally integrated magnetic flux in coronal holes above a specific latitude
\begin{equation}
  \Phi^{\rm CH}({\lambda_0})
  = R_\odot^2 \left|
    \int_{0}^{2\pi}\int_{\lambda_0}^{\pi/2}
    M_{\rm CH} B_r \cos\lambda d\lambda d\phi
  \right|
  + R_\odot^2 \left|
    \int_{0}^{2\pi}\int_{-\pi/2}^{-\lambda_0}
    M_{\rm CH} B_r \cos\lambda d\lambda d\phi
  \right|
\end{equation}
as a rough measure of the open magnetic flux.
Here, \(M_{\rm CH}\) is the mask function for coronal holes, which is defined as \(1\) if the AIA 193 intensity is lower than a threshold value and \(0\) otherwise. The magnetic field strength \(B_r\) is derived from the HMI synoptic maps. The threshold latitude \(\lambda_0\) is chosen to be \(10^\circ\), \(30^\circ\) or \(50^\circ\).
This measure is introduced to investigate the effect of the detection parameters on the coronal hole magnetic flux.
As a similar measure, we also defined the area of coronal holes above a threshold latitude \(\lambda_0\) as
\begin{equation}
  A^{\rm CH}({\lambda_0})
  = R_\odot^2 \left|
    \int_{0}^{2\pi}\int_{\lambda_0}^{\pi/2}
    M_{\rm CH} \cos\lambda d\lambda d\phi
  \right|
  + R_\odot^2 \left|
    \int_{0}^{2\pi}\int_{-\pi/2}^{-\lambda_0}
    M_{\rm CH} \cos\lambda d\lambda d\phi
  \right|.
\end{equation}

\begin{figure}[ht!]
  \epsscale{0.8}
  \plotone{{\figdir}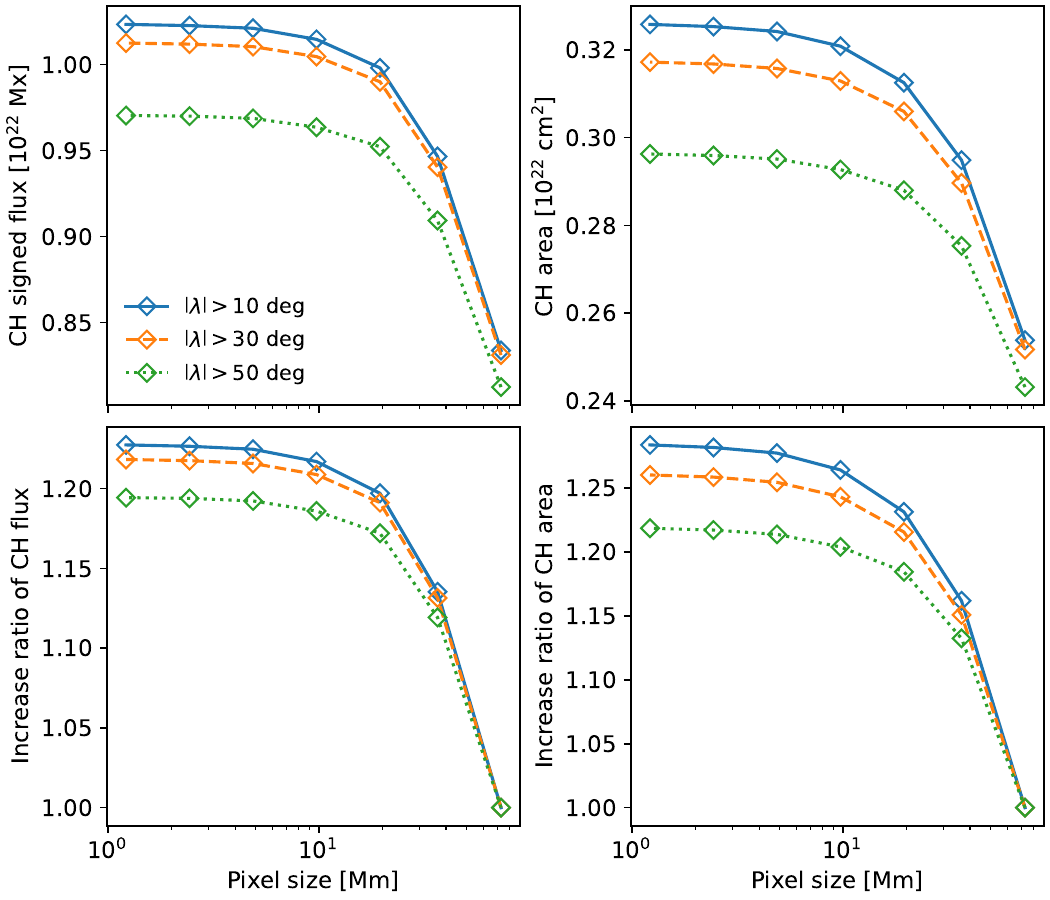}
  \caption{
    Dependence of the magnetic flux \(\Phi^{\rm CH}\) (left) and area \(A^{\rm CH}\) (right) in coronal holes on the spatial resolution.
    The top panels show the {total signed} magnetic flux and area of coronal holes above \(10^\circ\), \(30^\circ\), and \(50^\circ\).
    The bottom panels show the increase of magnetic flux and area relative to the pixel size of \(72\) Mm near the equator.
    The threshold value of AIA 193 intensity is set to \(20\) DN s\(^{-1}\) pix\(^{-1}\).
  }\label{fig:obs:deps_bin}
\end{figure}

Figure~\ref{fig:obs:deps_bin} shows the dependence of the magnetic flux \(\Phi^{\rm CH}\) and area \(A^{\rm CH}\) in coronal holes on the spatial resolution. Higher spatial resolution leads to increased flux estimates by several tens of percents. The effect of spatial resolution is slightly larger in regions closer to the equator. The increase rate of the coronal hole magnetic flux is similar to the coronal hole area.

\begin{figure}[ht!]
  \epsscale{0.8}
  \plotone{{\figdir}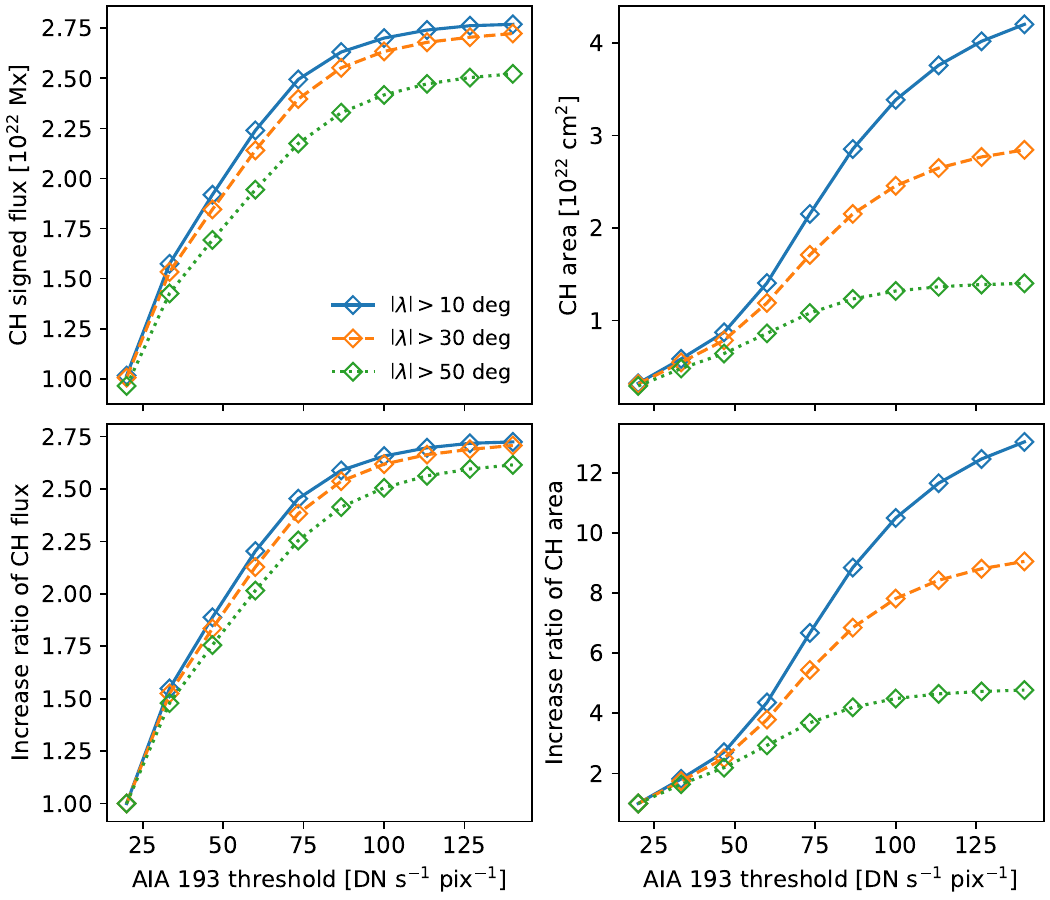}
  \caption{
    Similar to Fig.~\ref{fig:obs:deps_bin}, but showing the dependence on the brightness threshold of AIA 193 intensity.
    The pixel size is set to \(7.2\) Mm near the equator.
  }\label{fig:obs:deps_aia}
\end{figure}

Left panels in Fig.~\ref{fig:obs:deps_aia} show the dependence of the magnetic flux \(\Phi^{\rm CH}\) in coronal holes on the brightness threshold of AIA 193 intensity.
Changing the AIA 193 intensity threshold leads to an increase by a factor of \(2\) in derived coronal hole open flux.
This increase is large enough to explain the difference between the open flux estimations from in-situ and remote observations \citep{linker2017open}.
The result suggests a potential contribution from hidden open-field regions to the total open magnetic flux.

{
Right panels in Fig.~\ref{fig:obs:deps_aia} show that the increase in coronal hole area \(A^{\rm CH}\) is much larger than the increase in coronal hole magnetic flux, especially at lower latitudes.
The relative (per-area) effect of ``missing'' coronal holes on the total open magnetic flux is smaller at lower latitudes.
Using a higher brightness threshold of AIA 193 intensity helps reduce the effect of undetected coronal holes on the total open magnetic flux especially at higher latitudes.
}

\subsection{Comparison with in-situ observation of open magnetic flux}

\begin{figure}[ht!]
  \epsscale{0.6}
  \plotone{{\figdir}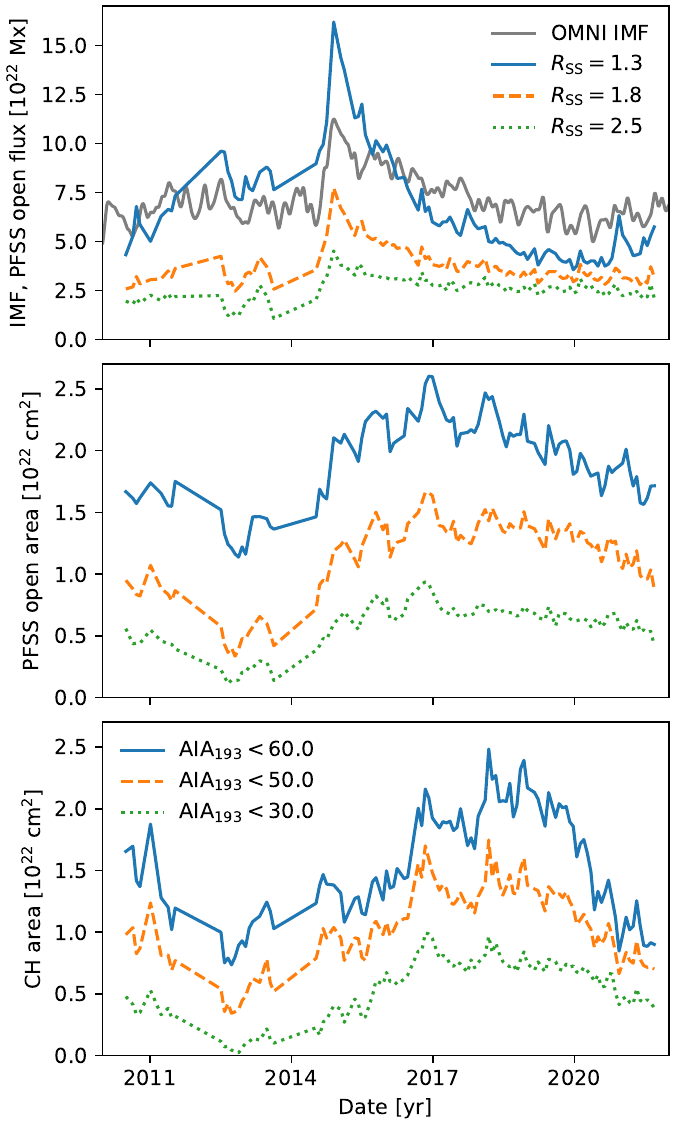}
  \caption{
    Time variation of open magnetic flux (top), total area of open-field regions (middle), and total area of coronal holes (bottom).
    The open magnetic flux in the top panel is estimated from the in-situ observation of interplanetary magnetic field in the OMNI data (gray solid line) and the PFSS model with a source surface height \(R_{\rm SS}\) of \(1.3 R_\odot\) (blue solid line), \(1.8 R_\odot\) (orange dashed line), and \(2.5 R_\odot\) (green dotted line).
    The total area of open-field regions (middle) is also estimated from the PFSS model.
    The total area of coronal holes (bottom) are estimated from the AIA 193 intensity with threshold values of \(60\) (blue solid line), \(50\) (orange dashed line), and \(30\) (green dotted line) DN s\(^{-1}\) pix\(^{-1}\).
  }\label{fig:obs:open_flux}
\end{figure}

Finally, we compare the open magnetic flux estimated from the PFSS model with the in-situ observations of interplanetary magnetic field in the OMNI data.
If the open-field region corresponds to the location of coronal holes with higher threshold value of AIA 193 intensity, the open magnetic flux can be approximated by the PFSS model with lower source surface height \(R_{\rm SS}\).
The top panel of Fig.~\ref{fig:obs:open_flux} shows the time variation of open magnetic flux estimated from the in-situ observation and the PFSS model with different source surface heights \(R_{\rm SS}\) of \(1.3 R_\odot\), \(1.8 R_\odot\), and \(2.5 R_\odot\).
The open magnetic flux estimated by the PFSS model with \(R_{\rm SS} = 2.5 R_\odot\) is several times smaller than the OMNI flux, as previously reported \citep{linker2017open}.
To approximate the magnitude of open magnetic flux estimated from the OMNI data, the PFSS model with \(R_{\rm SS} = 1.3 R_\odot\) is required. However, the PFSS open flux with \(R_{\rm SS} = 1.3 R_\odot\) is larger than the OMNI flux near the cycle maximum and smaller than the OMNI flux near the cycle minimum. This has been suggested as the time variation of the optimal source surface height \citep{levine1982open,lee2011coronal,huang2024adjusting}.
The similar temporal evolution of the total area of open-field regions (middle panel) and coronal holes (bottom panel) in Fig.~\ref{fig:obs:open_flux} supports the link between them over the solar cycle, consistent with our finding that higher brightness thresholds for coronal hole detection correspond to open-field regions derived using lower \(R_{\rm SS}\) values.

\begin{figure}[ht!]
  \epsscale{0.7}
  \plotone{{\figdir}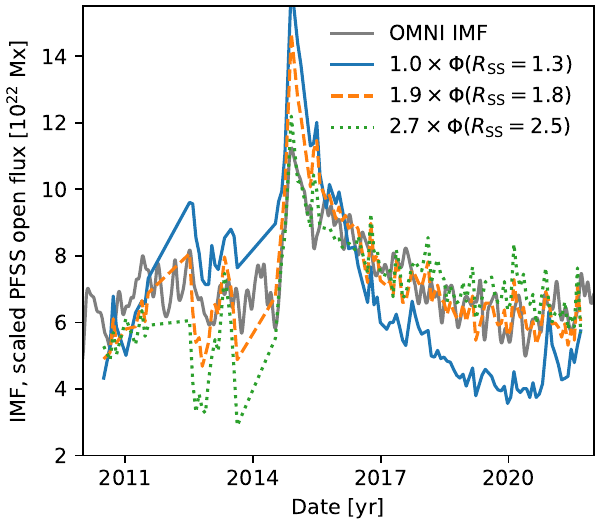}
  \caption{
    Similar to top panel of Fig.~\ref{fig:obs:open_flux}, but showing the open magnetic flux estimated by the PFSS model linearly scaled to match the open magnetic flux estimated from the OMNI IMF data.
    The scaling factors are \(1.0\) for \(R_{\rm SS} = 1.3 R_\odot\),
    \(1.9\) for \(R_{\rm SS} = 1.8 R_\odot\),
    and \(2.7\) for \(R_{\rm SS} = 2.5 R_\odot\), respectively.
  }\label{fig:obs:open_flux_scaled}
\end{figure}

Figure~\ref{fig:obs:open_flux_scaled} shows the open magnetic flux estimated by the PFSS model linearly scaled to match the open magnetic flux estimated from the OMNI IMF data. If we assume that the source surface height is constant in time, the time variation of open magnetic flux seems to be overestimated in the PFSS model with \(R_{\rm SS} = 1.3 R_\odot\). In this point of view, the source surface height \(R_{\rm SS}\) of \(1.8 R_\odot\) and \(2.5 R_\odot\) are preferable.
Recent studies \citep{milic2024spatial,sinjan2024magnetograms} report the existence of systematic biases in the magnetic field measurement on the photosphere especially near the polar region (near the limb of solar disk).
Thus, we speculate that the difference between the in-situ and remote observation of open magnetic flux \citep{linker2017open} could be influenced not only by the source surface height but also by the systematic bias in the photospheric magnetic field measurements.

\section{Summary and Discussion\label{sec:summary}}

In this study, we investigated the brightness of the magnetically open corona based on the three-dimensional radiative MHD simulations and the remote-sensing observation from the SDO HMI and AIA. Our findings from the numerical simulations are summarized as follows:
\begin{itemize}
  \item The spatially averaged brightness of magnetically open corona can reach the level of quiet regions. The simulated corona tends to be darker when the small-scale dynamo is weaker.
  \item Brightness of magnetically open corona is mainly determined by the total energy input into the corona rather than the difference in the dissipation process in the corona.
  \item The location of open-field region can be approximated by the brightness of corona (e.g. synthesized AIA 193 intensity), but the accuracy depends on the value of {brightness} threshold and spatial resolution.
\end{itemize}
Motivated by these simulation results, we utilized observational data and the PFSS model to test the hypothesis that bright coronal regions, similar to quiet regions, can contribute significantly to the open magnetic flux. Our findings from the observational analysis are summarized as follows:
\begin{itemize}
  \item The spatial location of coronal holes and open-field regions are closely related.
  \item The local magnetic field strength (as a measure of small-scale dynamo strength) cannot be a good indicator of open-field regions.
  \item The total open magnetic flux in coronal holes strongly depends on the threshold value of the coronal brightness (AIA 193 intensity in this study) used for the coronal hole detection.
  \item Spatially small or bright open-field regions might be missed in the open flux estimation based on coronal holes.
\end{itemize}

\subsection{What makes coronal hole dark?}

~\cite{pneuman1973solar} proposed that the coronal holes are dark because they are open to space.
As the energy flux (mainly Poynting flux due to Alfv\'en waves) is transported along the magnetic field lines in the corona, magnetically open regions allows a larger energy escape into space than closed regions.
Escaped energy is not available for heating the corona, but instead accelerates the solar wind.
Our simulations confirmed this idea, showing that the magnetically open numerical corona tends to be darker than the closed regions.

Coronal holes are more magnetically unipolar than the surrounding quiet regions \citep[e.g.,][]{harvey1979coronal,levine1982open,abramenko2009parameters}.~\cite{abramenko2006rate} and \cite{hagenaar2008dependence} showed that coronal holes exhibit a lower rate of flux emergence, thus leading to a lower energy input into the corona.
This result is qualitatively consistent with our simulation results, which show that the dissipation per unit energy input decreases with larger magnetic flux imbalance.
However, we emphasize that the variation of energy input itself is much more significant than the parameter dependence of dissipation per unit energy input.
This strong dependence of energy input on the photospheric magnetic energy is also observed in our solar wind simulation \citep{iijima2023comprehensive}, suggesting the contribution of interchange magnetic reconnection.

{
Thus, we suggest two physical conditions for the formation of coronal holes: (1) the magnetic field is connected to space, and (2) the photospheric magnetic energy is relatively weak.
Regarding the second point, we are aware that internetwork magnetic energy in quiet regions is generally uniform \citep[e.g.,][]{lites2011hinode,buehler2013quiet,korpi2022solar}.
Nevertheless, we stress this point for three reasons: (i) in our simulations the realized corona depends sensitively on the prescribed dynamo strength, (ii) the magnetic energy criterion may be important when considering solar-like stars, and (iii) the dispersion of the photospheric magnetic energy is not negligible.
For instance, Fig.~6 of \citet{korpi2022solar} shows that the dispersion in quiet regions is typically several tens of percent, depending on the size of the horizontal averaging region.
Brighter coronae near active regions also exhibit higher magnetic energy.
Additionally, the removal of active regions in the studies of quiet-Sun magnetic field may reduce the dispersion of inferred surface magnetic energy.
While further work is certainly required, these considerations suggest that magnetic energy remains an important factor for understanding coronal hole formation.
}

\subsection{Possibility of ``hidden'' open-field regions}

From the numerical simulations and observations, we suggest that the large amount of open-field regions are hidden or not categorized as coronal holes. Such bright but open-field regions may be already observed as coronal upflows in the spectroscopic observation of the Sun. For example, the active region upflows \citep{sakao2007continuous,harra2008outflows} have been discussed as a potential source of the slow solar wind \citep{brooks2015full}. Ubiquitous coronal upflows are also found in coronal holes \citep{xia2004network}. For more information on the observation of coronal upflows, we refer the reader for the review by \cite{tian2021upflows}.
As we focused on the coronal brightness in this study, it remains unclear whether spectroscopic observations can suggest better indicators of open-field regions.
This possibility should be investigated in future studies.
The time-steady modelings of coronal magnetic field like the potential field, nonlinear force-free, or magnetohydrostatic models may not be possible to reproduce these intermittent nonlinear structures (sometimes observed as coronal jets). Recent studies \citep{bale2023interchange,iijima2023comprehensive} have suggested the interchange reconnection plays a strong impact in the solar wind formation. For better understanding of the coronal magnetic field structure, time-dependent high-resolution models are preferable.

\subsection{Implications for Source Surface Heights\label{sec:summary:ss_height}}

The source surface in the PFSS model can be considered as an approximate measure of where the solar wind is sufficiently accelerated and the magnetic field is decoupled from the solar surface.
The optimal height of the source surface has been discussed, but remains unclear \citep{levine1982open,lee2011coronal,yoshida2023component,huang2024adjusting,tokumaru2024coronal}.
Some studies have explored the possibility of non-spherically symmetric shape of the source surface \citep{kruse2020elliptic}.
Lower heights of source surface may result from the rapid acceleration of solar wind near the Sun \citep{rice2021global}.
We also suggest that the source surface height may be lower than the commonly used value of \(R_{\rm SS} = 2.5 R_{\odot}\) \citep{schatten1969model,hoeksema1983structure} in this study.
This diversity in the source surface is a natural consequence of the strong simplification of the PFSS model.
We hope that the future development of global coronal MHD models will help resolve these issues.

In this study, we suggest that lowering the source surface height is a possible solution for the ``open flux problem'' \citep{linker2017open}.
However, we also confirmed the good correspondence between open-field regions and coronal holes, as long as the threshold value for the coronal hole detection is appropriately set.
{Nonetheless, determining an appropriate threshold is itself nontrivial and can introduce additional uncertainty.}
Additionally, we found that the source surface height of \(1.3 R_\odot\) may be too low, as the {time variability} of total open magnetic flux {exceeds that indicated by} in-situ observations (Fig.\ref{fig:obs:open_flux_scaled}).
Our findings suggest that the criteria for identifying coronal holes as proxies for open-field regions warrant reconsideration, particularly because this proxy is widely used to validate coronal magnetic-field extrapolation models.
The estimated open magnetic flux is likely affected by both inaccuracies in coronal-field modeling and systematic biases in photospheric magnetic-field measurements \citep{milic2024spatial,sinjan2024magnetograms}.

\begin{acknowledgments}
  We thank D. Shiota and M. Rempel for fruitful discussions.
  We also thank the anonymous referee for the constructive comments that helped to improve the manuscript.
  This work used computational resources of supercomputer Fugaku provided by the RIKEN Center for Computational Science through the HPCI System Research Project (Project ID: hp220147).
  Data analysis was performed in the Center for Integrated Data Science, Institute for Space-Earth Environmental Research, Nagoya University.
  The OMNI data were obtained from the GSFC/SPDF OMNIWeb interface at \url{https://omniweb.gsfc.nasa.gov}.
  This work was supported by JSPS KAKENHI grant No. JP21H01124 and JP23K13144, MEXT as ``Program for Promoting Researches on the Supercomputer Fugaku'' (Elucidation of the Sun-Earth environment using simulations and AI; Grant Number JPMXP1020230504), and the Program for Promoting the Enhancement of Research Universities, as a young researcher unit for the advancement of new and undeveloped fields, Nagoya University.
\end{acknowledgments}

%

\appendix

\section{Correction for limb brightening of AIA synoptic maps}
\label{appendix:lbcc}

\begin{figure}[ht!]
  \plotone{{\figdir}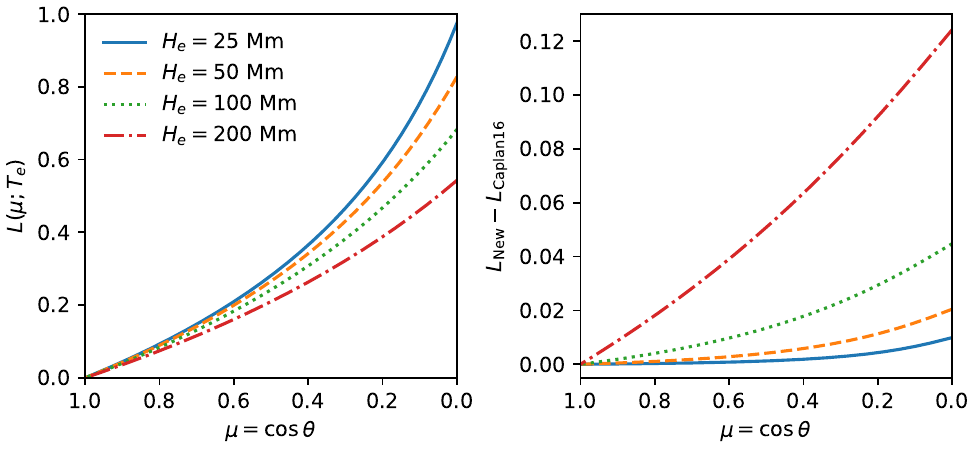}
  \caption{
    Theoretical limb brightening correction curves (LBCCs) used to correct the limb brightening of AIA 193 \AA\ intensity in this study.
    Left panel shows the LBCCs for different electron density scale heights.
    Right panel shows the difference from the LBCC suggested by~\cite{caplan2016synchronic}.
    }\label{fig:app:lbcc}
\end{figure}

The AIA 193 \AA\ intensity maps are affected by the limb brightening, which is caused by the line-of-sight superposition of the coronal plasma. The limb brightening effect should be corrected to accurately detect the coronal hole based on the spatially and temporally uniform threshold intensity.
We followed the method suggested by \cite{caplan2016synchronic}. Assuming the plasma stratification in the corona, they derived theoretical and empirical models of the limb brightening.
In this study, we used a version with a constant gas pressure scale height, with minor modifications.

We assume the following exponential vertical profile of the electron density
\begin{equation}
  n_e(r) = n_0 \exp\left[-\frac{r-R_0}{H_e}\right],
  \label{eq:lbcc:density}
\end{equation}
where \(n_0\) is the electron density at the base of the corona, \(R_0\) is the radial distance from the solar center to the base of the corona, and \(H_e\) is the scale height, respectively.
Following \cite{caplan2016synchronic}, we set \(R_0 = 1.01 R_\odot\) in this study.

The difference from the original model is that we assume a constant gas pressure scale height. Although~\cite{caplan2016synchronic} used the scale height consistent with the inverse-square law of gravity, we choose the above relation to simplify the calculation. The true coronal plasma is not in hydrostatic equilibrium, isothermal, nor spherically symmetric. Additionally, the existence of closed coronal loops confines the plasma locally reducing the effective scale height. Thus, we expect that the above exponential profile (Eq.\ref{eq:lbcc:density}) is reasonable.

Integrating the spherically symmetric vertical profile of Eq.\ref{eq:lbcc:density}, the {radiative} intensity is given by
\begin{equation}
  F(\mu; T)
  = \int_0^\infty G(T)n_e^2 ds
  = G(T)n_1^2 \int_0^\infty \exp\left[-\frac{2\sqrt{s^2 + 2 s \mu R_0 + R_0^2}}{H_e}\right] ds,
\end{equation}
where \(s\) is the distance along the line-of-sight, \(\mu\) is the cosine of angle from disk center (\(\mu = 1\) corresponds to the disk center), \(T\) is the gas temperature (assumed to be constant along the line-of-sight), \(G(T)\) is the response function of AIA 193 \AA\ intensity, and \(n_1 = n_0 \exp(-R_0/H_e)\) is a constant.

As a ratio between theoretical intensities \(F(\mu; T)\) at different line-of-sight inclinations, we can derive the limb brightening correction curve (LBCC) as
\begin{equation}
  L(\mu; T) = \log_{10}\frac{F(\mu; T)}{F(\mu=1; T)}.
\end{equation}
Left panel in Fig.~\ref{fig:app:lbcc} shows the numerically evaluated LBCCs for the AIA 193 \AA\ intensity. Evidently, the theoretical intensity increses with the line-of-sight inclination and with a smaller density scale height.
By dividing the observed intensity by \(10^{L(\mu; T)}\), we can remove the limb brightening effect.
Right panel in Fig.~\ref{fig:app:lbcc} shows the difference from the LBCC suggested by~\cite{caplan2016synchronic}. The difference is less than \(0.02\) for the density scale height of \(H_e = 50\) Mm.
In this study, we used the LBCC with \(H_e = 50\) Mm {as a reasonable value for the coronal plasma} to correct the limb brightening effect in the AIA 193 \AA\ intensity maps.



\end{document}